\documentclass[submitting]{nst}

\usepackage{subfigure,dcolumn}
\usepackage{epstopdf}
\usepackage{mhchem}
\usepackage{upgreek}
\usepackage{float}
\usepackage{graphicx}
\usepackage{hyperref}
\usepackage{color}

\usepackage{xcolor}
\usepackage{ulem}

\newcommand {\rootsNN}  {\ensuremath{\sqrt{s_{_{\rm{NN}}}}}}

\begin{document}

\title{Event plane determination from Zero Degree Calorimeter at the Cooling-Storage-Ring External-target Experiment}
\thanks{This work is supported in part by the National Key Research and Development Program of China under contract Nos. 2022YFA1604900 and 2020YFE0202002; 
the National Natural Science Foundation of China (NSFC) under contract Nos. 12175084, 11890710 (11890711) and 11927901; the Strategic Priority Research Program of Chinese Academy of Sciences, Grant No. XDB34030000; 
and the Fundamental Research Funds for the Central Universities (CCNU220N003).}

\author{Li-Ke Liu}
\affiliation{Key Laboratory of Quark \& Lepton Physics (MOE) and Institute of Particle Physics, 
Central China Normal University, Wuhan 430079, China}

\author{Hua Pei}
\affiliation{Key Laboratory of Quark \& Lepton Physics (MOE) and Institute of Particle Physics, 
Central China Normal University, Wuhan 430079, China}

\author{Yaping Wang}
\affiliation{Key Laboratory of Quark \& Lepton Physics (MOE) and Institute of Particle Physics, 
Central China Normal University, Wuhan 430079, China}

\author{Biao Zhang}
\affiliation{Key Laboratory of Quark \& Lepton Physics (MOE) and Institute of Particle Physics, 
Central China Normal University, Wuhan 430079, China}

\author{Nu Xu}
\affiliation{Key Laboratory of Quark \& Lepton Physics (MOE) and Institute of Particle Physics, 
Central China Normal University, Wuhan 430079, China}

\author{Shusu Shi}
\email[Corresponding author, ]{shiss@mail.ccnu.edu.cn}
\affiliation{Key Laboratory of Quark \& Lepton Physics (MOE) and Institute of Particle Physics, 
Central China Normal University, Wuhan 430079, China}

\begin{abstract}
The Cooling-Storage-Ring External-target Experiment (CSR-CEE) is a spectrometer to study the nature of nuclear matter created 
in heavy ion collision at {\rootsNN} =  2.1-2.4 GeV, aiming to reveal Quantum Chromodynamics (QCD) phase structure in the high-baryon density region. 
Collective flow is regarded as an effective probe for studying the properties of the medium in high-energy nuclear collisions. One of the main functions of the Zero-Degree Calorimeter (ZDC), a sub-detector system in CEE, 
is to determine the reaction-plane in heavy ion collisions, which is crucial for the measurements of collective flow and other reaction plane related analysis.
In this paper, we illustrate the procedures of event plane determination from ZDC. 
Finally, predictions of the rapidity dependence of directed and elliptic flow for $p$, $d$, $t$, $^3$He and $^4$He, from 2.1 GeV U+U collisions of IQMD model calculations, are presented.
\end{abstract}


\maketitle
\nolinenumbers

\section{Introduction}\label{sec.I}

At sufficiently high temperature and/or high density, Quantum Chromodynamics (QCD) predicts a phase transition from hadronic matter to deconfined partonic matter~\cite{Braun-Munzinger:2007edi}.
Results from top RHIC and LHC energies indicate 
a new form of matter with small viscosity and high temperature, Quark-Gluon Plasma (QGP), has been produced~\cite{BRAHMS:2004adc, PHOBOS:2004zne, PHENIX:2004vcz, STAR:2005gfr, Bazavov:2011nk, Fukushima:2013rx}. 
Lattice QCD calculations predict that, the phase transition from hadronic matter to the QGP phase is a smooth crossover at vanishing baryon chemical potential ($\mu_{B}$) region~\cite{Aoki:2006we}. A first-order phase transition is expected at a finite baryon chemical potential region, revealing the phase structure of QCD is a major research goal in the field of medium and high energy heavy-ion collisions~\cite{phaseTran:2010, Bzdak:2019pkr, Luo:2020pef, Huang:2023ibh}.

The Cooling-Storage-Ring External-target Experiment (CEE) is a spectrometer to study the properties of nuclear matter at 2.1-2.4 GeV energy region in the center-of-mass frame~\cite{Lu:2016htm}.
Its main function is to achieve near-full-space measurements of charged particle in heavy-ion collisions, and to provide experimental data for the study of important scientific problems such as spin- and isospin-related nuclear forces, nuclear matter equations of state and QCD phase structure in high baryon number density~\cite{Horowitz:2014bja, Zhang:2009ba, Andronic:2009gj}. It will offer valuable research opportunities for QCD phase diagram studies in the low-temperature and high-baryon density region.

Event anisotropy of final state particles relative to the reaction plane in momentum space, which is also known as collective flow~\cite{Voloshin:2008dg}, is an important observable to study the medium properties created in heavy-ion collisions.
The flow coefficients, such as directed flow $v_1$ and elliptic flow $v_2$, are characterized by the harmonic coefficients in the Fourier expansion of the azimuthal distribution of final particles with respect to reaction plane. The driving force of collective flow is from the initial anisotropy in coordinate space in heavy-ion collisions. It diminishes rapidly as a function of time, known as the self-quenching effect. Thus collective flow is sensitive to the details of the expansion of the nuclear matter during the early collision stage.
Directed flow $v_1$ is predicted to be sensitive to the effective equation-of-state (EoS)~\cite{Bass:1998ca, Steinheimer:2022gqb, Oliinychenko:2022uvy}. Elliptic flow $v_2$ is sensitive to the the constituent interactions and degree of freedom~\cite{STAR:2015gge, STAR:2017kkh, Shi:2016elm}.
The CEE experiment can provide measurements of collective flow in heavy ion collisions 
at {\rootsNN} = 2.1-2.4 GeV. It will help us to study the medium properties and further search for the possible QCD phase transition signals~\cite{Yasushi:JAM_update2021, Yasushi:Lambdav1_2022, Lan:model2022}. 
One of the main functions of the Zero Degree Calorimeter (ZDC), a sub-detector of the CEE, is designed to determine the reaction plane in nucleus-nucleus collisions. The reconstructed reaction plane (usually called event plane) is
crucial for many measurements, such as collective flow~\cite{STAR:v1_2014, STAR:2021yiu, Nara:2022ixo}, azimuthal HBT~\cite{STAR:HBT2015}, CME related observables~\cite{CME:2008, STAR:CME200, Zhao:2022grq, Chen:2023jhx} and so on.

In this paper, we introduce necessary acceptance corrections and calibrations on the event plane determination 
from CEE-ZDC. A prediction of collective flow from a typical CEE energy ({\rootsNN} = 2.1 GeV) based on Isospin dependent Quantum Molecular Dynamics (IQMD)~\cite{Hartnack:1997ez} model is also shown at the end.

\section{CEE-ZDC}\label{sec.II}
\begin{figure}[htbp]
  \subfigure[] {
   \label{Fig:1a}     
  \includegraphics[width=0.51\columnwidth]{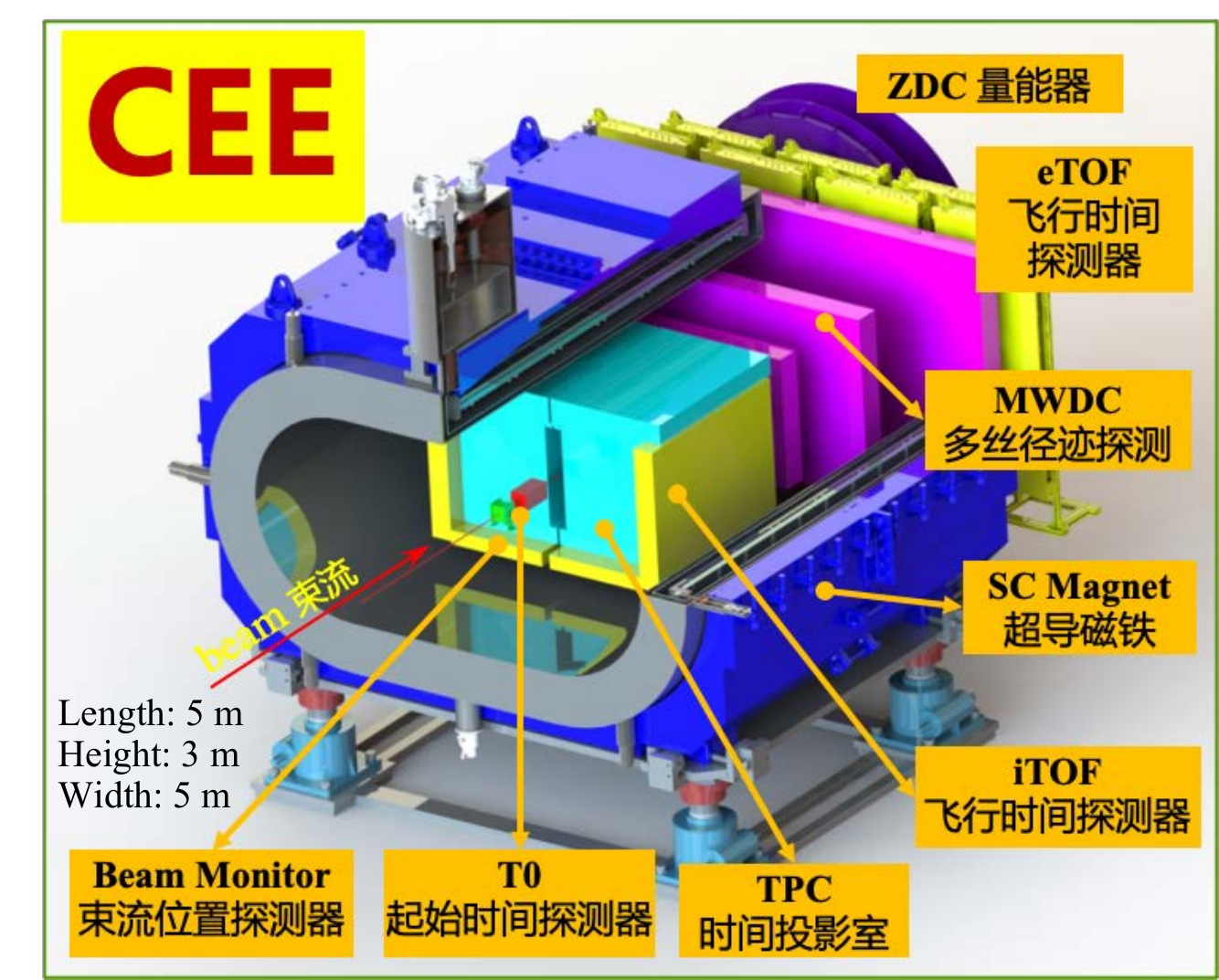}
  }
  \subfigure[] { 
  \label{Fig:1b}     
  \includegraphics[width=0.4\columnwidth]{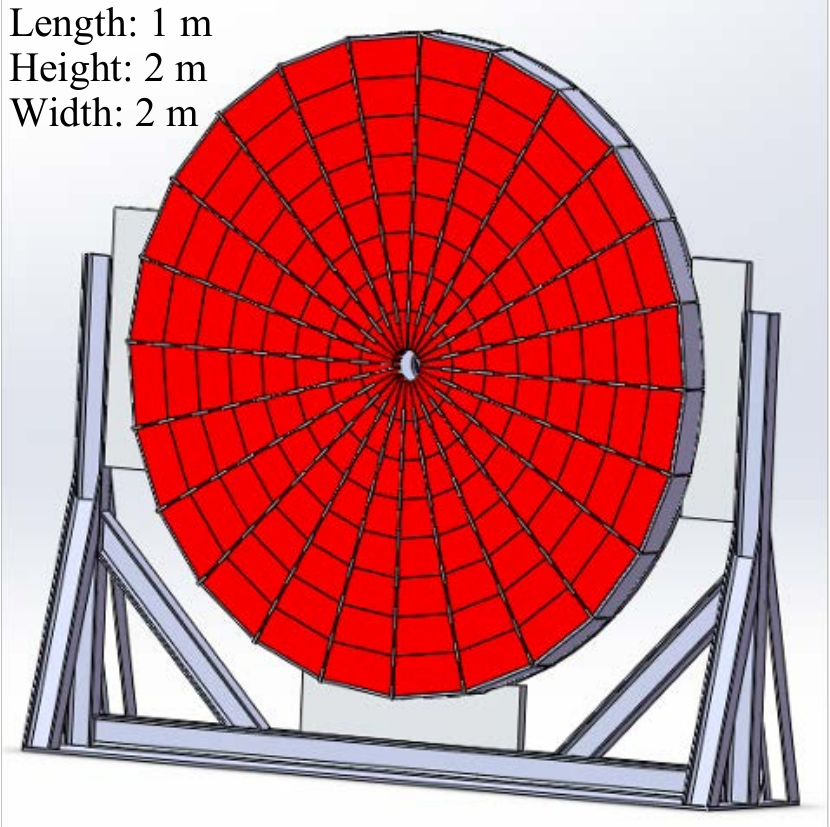}
  } 
  \caption{(a) The sketch of Cooling-Storage-Ring External-target Experiment (CEE) detector. (b) The sketch of ZDC detector.}     
  \label{Fig:CEE_ZDC}
\end{figure}

Figure~\ref{Fig:CEE_ZDC}(a) shows the sketch of the CEE spectrometer. The detector subsystem consists of: superconducting dipole magnet used to deflect charged particles; Silicon Pixel positioning detector (SiPiX, Beam Monitor) to measure the position, time of the incident beam, and primary collision vertex~\cite{CEE:BeamMonitor2022}; 
Time Projection Chamber (TPC) to reconstruct the particle trajectory and identify particles~\cite{Huang:2018dus}; 
Time-of-flight chamber (TOF) to extend particle identification to high momentum ($p > 2$ GeV/c), 
containing a start-time detector (T0)~\cite{CEE:T02019}, an inner time-of-flight detector (iTOF)~\cite{CEE:iTOF2022}, and an end-cap time-of-flight detector (eTOF)~\cite{CEE:eTOF2020}; 
Multi-Wire Drift Chambers (MWDC) is designed to track the charged particles at forward rapidity, and also participate in the particle identification via momentum measurement \cite{CEE:MWDC}; 
ZDC to measure the pattern (deposited energy and incident position) of forward-going charged particles emitted from nucleus-nucleus collisions~\cite{CEE:ZDC2021}.

The ZDC is proposed to be installed behind all other sub-detectors. The beam direction is defined as positive $Z$-axis and the ZDC is located at $Z$ = 295-299 cm, facing the original incident beam direction. Its geometry is illustrated in Fig.~\ref{Fig:CEE_ZDC}(b). ZDC detector cross plane is a wheel with a radius $R$ from 5 to 100 cm and the vacuum pipe carrying the nuclear beam passes through the inner hole of ZDC wheel. It consists of 24 sectors which subtend 15 degrees in azimuth. Each sector is divided into 8 modules, which form 8 rings in the full ZDC plane. 
The sensitive volume of ZDC is composed by plastic scintillator, and the current design selects BC-408 material from Saint-Gobain~\cite{ZDC:Saint-Gobain}. The photons are produced inside the scintillator through deposited energy of incident particle, and then transport through a plastic light guide into the quartz window of a traditional PMT.
ZDC will cover the pseudo-rapidity range between 1.8 and 4.8, allowing the determination of the centrality and the event plane in the forward rapidity region, minimizing auto-correlations from middle rapidity analyses~\cite{Voloshin:2008dg, STAR:EPD2019}.

\section{Event plane determination from CEE-ZDC}\label{sec.III}
In the study of the event plane, the simulation input of  in $^{238}$U + $^{238}$U collisions at 500 MeV/u is from the IQMD generator~\cite{Hartnack:1997ez}.
The IQMD model is developed based on the Quantum Molecular Dynamics (QMD) model~\cite{Aichelin:1991xy} considering isospin effects.
The detector environment is simulated by GEANT4~\cite{Brun:1994aa}. One million IQMD simulated events are generated in the range of nuclear impact parameter $b$, which is the transverse distance of the projectile from the target nucleus, $0 < b < 10$ fm, with 0.1 million events for each $b$ interval of 1 fm.

The reaction plane in nucleus-nucleus collision is defined by the vector of the impact parameter and the beam direction. 
Since the impact parameter can not be directly measured in experiment, 
the reaction plane is estimated by standard event plane method~\cite{Poskanzer:1998yz, Voloshin:2008dg}. 
The first order harmonic event plane $\Psi_{1}$ is calculated by event flow vector $\vec{Q}_{1}$, 
\vspace{0.16cm}
\begin{equation}
  \vec{Q}_{1}=\left(\begin{array}{c}
    \sum_{i} w_{i} \sin \left(\phi_{i}\right) \\
    \sum_{i} w_{i} \cos \left(\phi_{i}\right)
    \end{array}\right)
    \quad \Psi_{1}=\tan ^{-1}\left(\frac{\sum_{i} w_{i} \sin \left(\phi_{i}\right)}
    {\sum_{i} w_{i} \cos \left(\phi_{i}\right)}\right) 
  \label{Eq:Psi_Q}
\end{equation}

where the sum goes over all particles used in the flow vector calculation. The quantities of $\phi_{i}$ is azimuth in the laboratory frame.  
The weight, $w_i$, is defined by the deposited energy $\Delta E$ of particle $i$ collected by ZDC detector. 
As it is related to the mass and transverse momentum $p_{\mathrm{T}}$ value of particle,
while the $p_{\mathrm{T}}$ weight is commonly applied in flow analysis to optimize the event plane resolution~\cite{Poskanzer:1998yz}.
The smearing effect of deposited energy is considered by Equ.~\ref{Eq:Energy_smear}
\begin{equation}
\begin{aligned}
& \Delta E= \Delta E^{\prime} * [1-\frac{1}{4}(\frac{L}{5.5})^{2} ], \quad h<8\\
& \Delta E= \Delta E^{\prime} * [1-\frac{1}{4}(\frac{L}{5.5})^{2} ]* [8+\frac{2}{3}(h-8)], \quad h\ge8
\end{aligned}
  \label{Eq:Energy_smear}
\end{equation}

where $L$ is the distance from the hit position to the geometric center of the sector,
$h$ is the charge number of the final particles.
The term $1-\frac{1}{4}(\frac{L}{5.5})^{2}$  is used to describe the deposited energy resolution at the edge of the sector, 
and the term $8+\frac{2}{3}(h-8)$ is used to simulate the saturation effect of the deposited energy resolution for the heavy nuclei ($h\ge8$)~\cite{Ding:2018lfn}.

Since finite multiplicity limits the estimation of the reaction plane, it brings a resolution factor $R$ which is defined by Equ.~\ref{Eq:Res}.
In this study, we focus on the first order harmonic event plane, as the $v_1$ is more significant than
higher orders flow in the range of collision energy covered by CEE. 
\begin{equation}
  R_{1}=\left\langle\cos \left(\Psi_{1, {EP}}-\Psi_{1, {RP}}\right) \right\rangle
  \label{Eq:Res}
\end{equation}

\begin{figure}[htbp]
  \centering   
  \includegraphics[width=0.49\columnwidth]{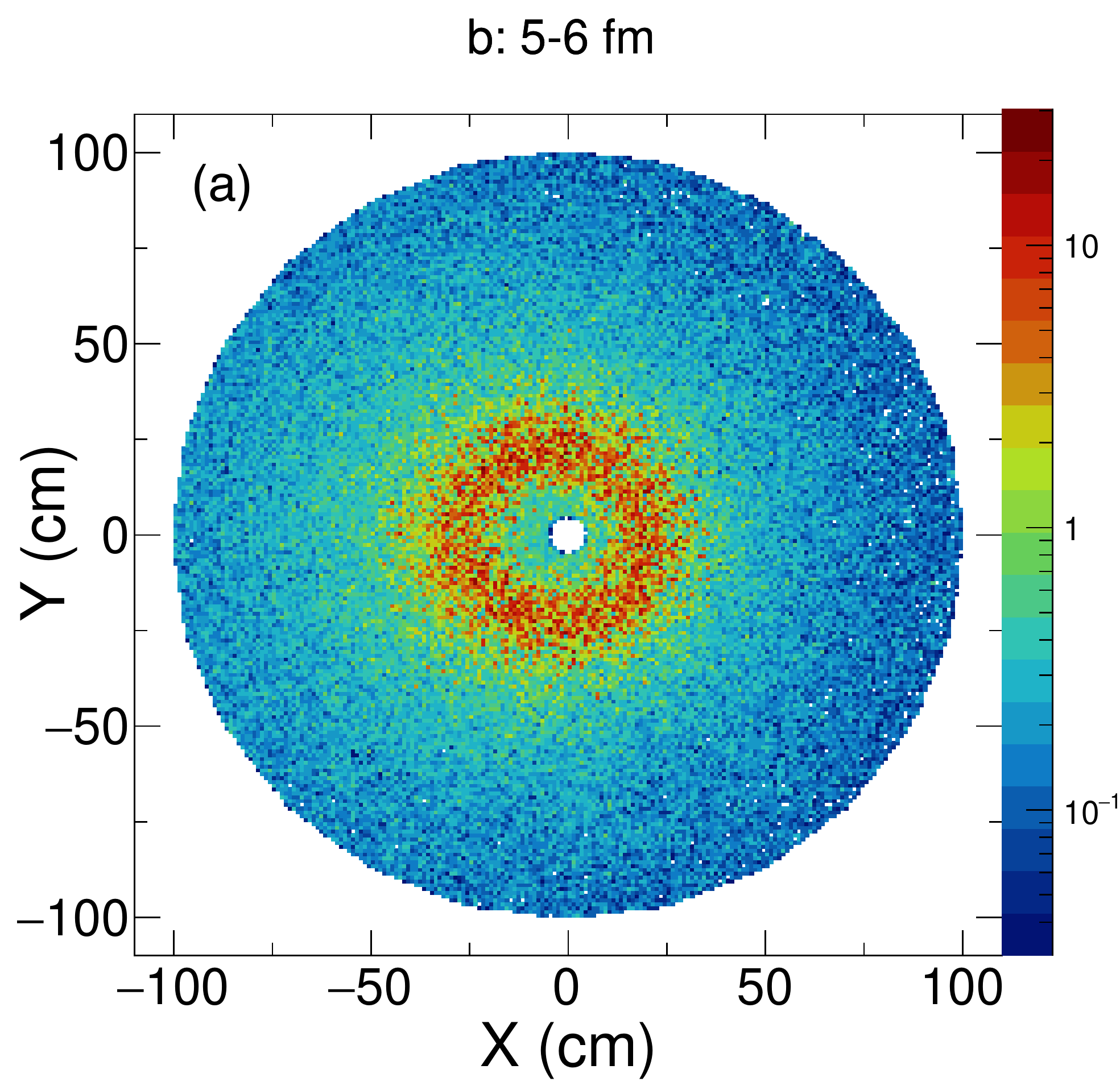}
  \includegraphics[width=0.49\columnwidth]{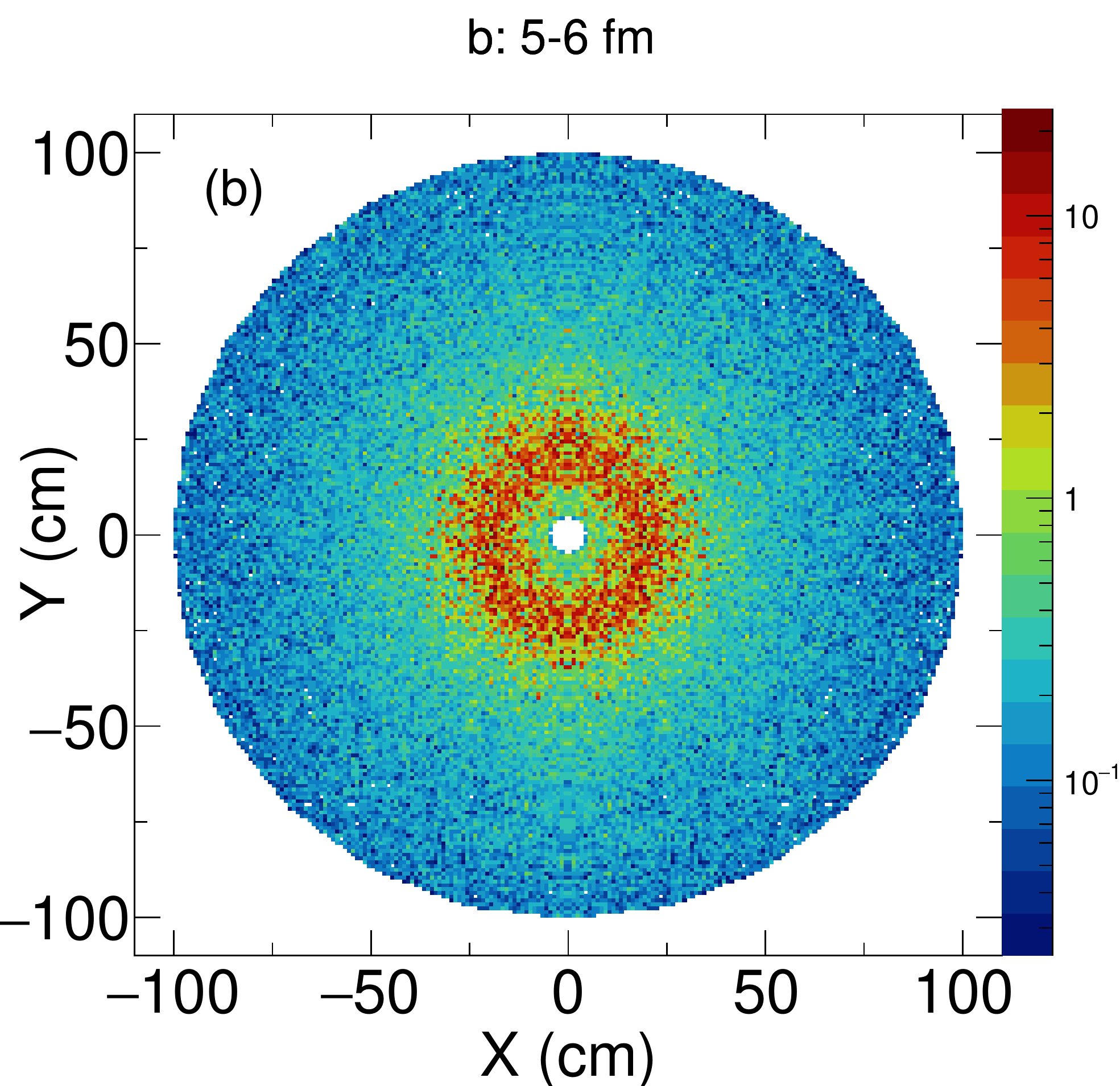}   
  \caption{(a) Hit distribution from ZDC with collision impact parameter  of $5 < b < 6 $ fm.
  (b) Hit distribution from ZDC after position weight correction with collision impact parameter of $5 < b < 6 $ fm.}     
  \label{Fig:Hit_distribution}
\end{figure}

The magnetic field direction is perpendicular to the beam direction at CEE, 
thus the charged particles of the final state are deflected by the magnetic field and hit one side of the ZDC detector more as Fig.~\ref{Fig:Hit_distribution}(a) shown. 
Due to the asymmetric ZDC acceptance, 
the reconstructed event plane angle is not isotropic in the laboratory frame,
but biased towards the $\pi$ azimuth.
The acceptance bias caused by magnetic field introduces an additional nonphysically anisotropy for the detected collision events, one should remove this effect as it distorts the event plane reconstruction.
Therefore, we introduce a position weight to calibrate the asymmetric acceptance.

The core idea of position weight is a correction to the asymmetric acceptance of ZDC
which is caused by the magnetic field.
Due to the deflection of charged particles in the magnetic field,
the left side of the ZDC detector receives more hits.
We assign a weight $P$ which is less than 1 to the hits on left side to correct this effect. 
The weight is calculated based on two dimensional $X-Y$ hit distribution as defined in Equ.~\ref{Eq:Position_weight}, is the ratio of the number of hits of the right side over the left side.
In addition, the deposited energy $\Delta E$ is also used as a weight when calculating the number of hits
as it is related to particle's mass.
One can observe the acceptance of ZDC is symmetric after applying the position weight as shown in Fig.~\ref{Fig:Hit_distribution}(b).

\begin{equation}
\begin{aligned}
&w_i=\Delta E \times P \\
&P=n(-x, y, \Delta E) / n(x, y, \Delta E), \quad x<0 \\
&P=1, \quad x>0
\end{aligned}
  \label{Eq:Position_weight}
\end{equation}

\begin{figure}[htb]
    \centering
    \includegraphics[width=2.4in,keepaspectratio]{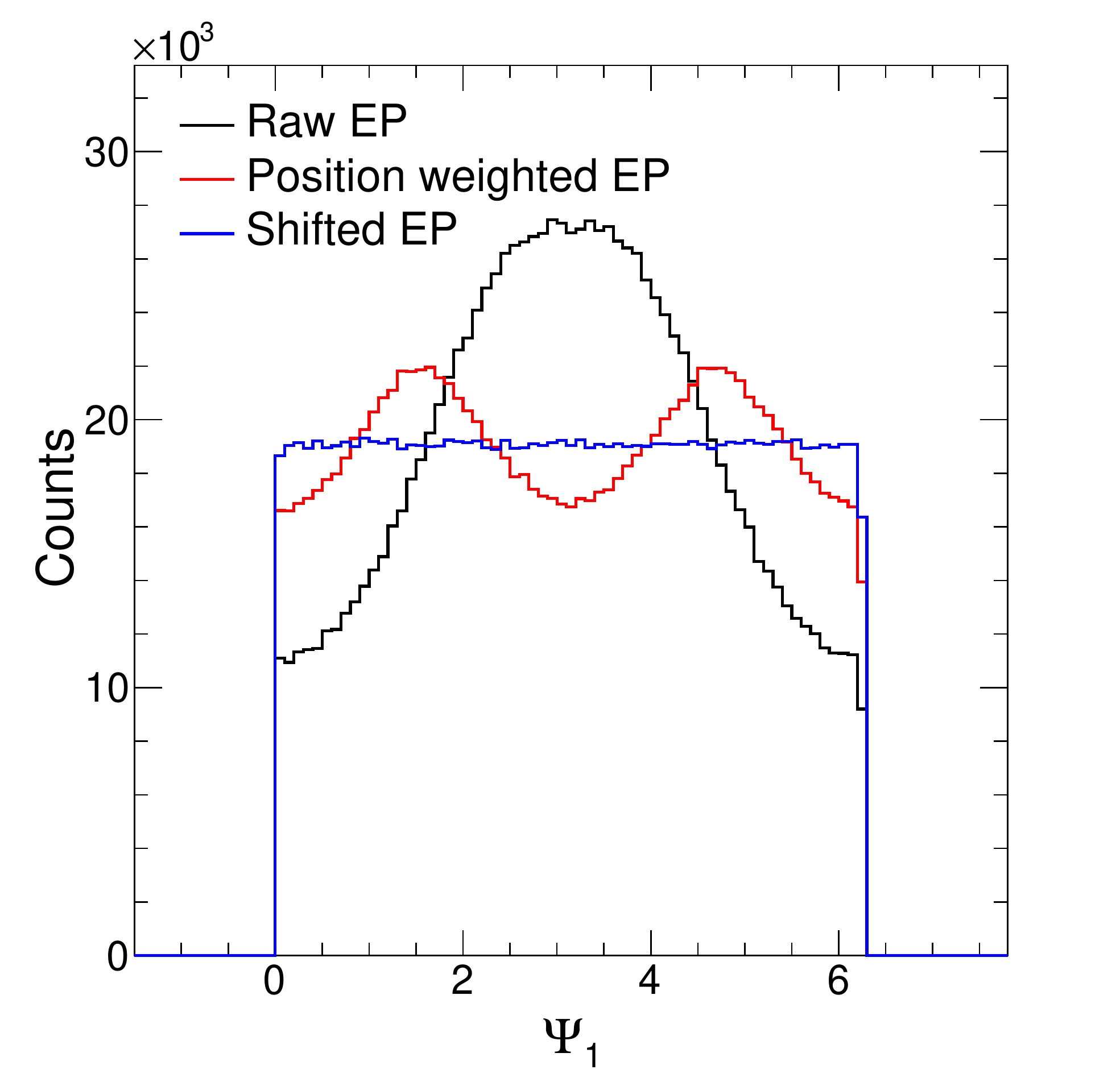}
    \caption{The event plane distributions before position weight correction (black line), after position weight correction (red line) and after position weight + shift correction (blue line).}
    \label{Fig:EP_dis}
\end{figure}

\begin{figure}[htb]
  \centering 
  \includegraphics[width=3.6in,keepaspectratio]{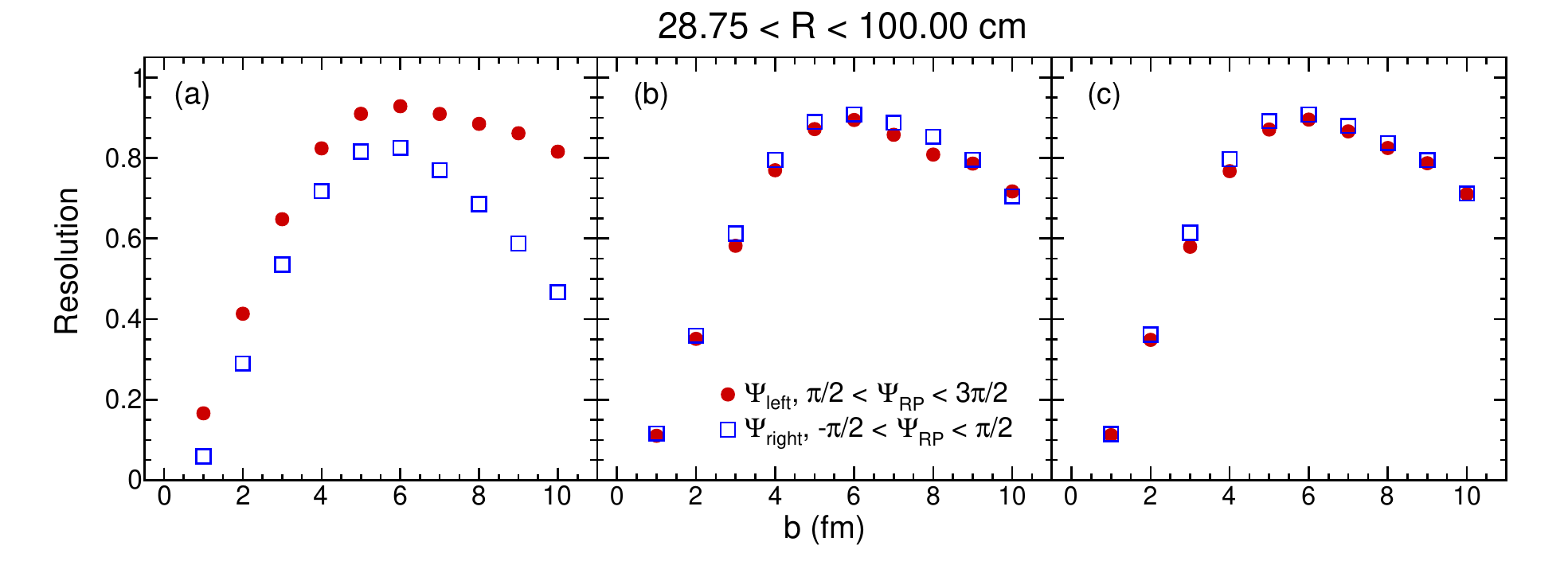}
  \caption{
  (a) The resolution of $1^{\rm st}$ order event plane as a function of impact parameter $b$ without position weight.
  (b) The resolution of $1^{\rm st}$ order event plane as a function of $b$ with position weight.
  (c) The resolution of $1^{\rm st}$ order event plane as a function of $b$ with position weight and shift correction.
  }     
  \label{Fig:whole_res}
\end{figure}

The black line in Fig.~\ref{Fig:EP_dis} shows the event plane distribution before the position weight correction.
With an ideal detector, the event plane distribution should be flat as the possible direction of impact parameter $b$ is random in the
$2\pi$ azimuth of transverse plane in the laboratory frame. 
It is not flat but peaked around $\Psi_{1} \sim \pi$ due to the asymmetric acceptance of ZDC as discussed above.
Correspondingly, one can see that the resolution difference between the left (the azimuth of reaction plane: $\pi/2$ to $3\pi/2$)  and right side ($-\pi/2$ to $\pi/2$) of ZDC is significant in Fig.~\ref{Fig:whole_res}(a).
After applying the position weight defined in  Equ.~\ref{Eq:Position_weight}, 
the unflatness of event plane is greatly reduced as shown by red line in Fig.~\ref{Fig:EP_dis}.
The resolution difference between the left and right side of ZDC is greatly reduced shown in Fig.~\ref{Fig:whole_res}(b). 
It indicates the position weight naturally corrects the acceptance asymmetry of ZDC.

The event plane distribution is not perfectly flat after the position weight~as shown in Fig.~\ref{Fig:EP_dis}.
As a consequence, the resolution difference from the left and right side of ZDC is still visible. 
Therefore, the shift method is further used to force event plane to be flat~\cite{Poskanzer:1998yz}. 
A shift angle $\Delta \Psi_{1}$ is applied to correct the event plane, 
and the $\Delta \Psi_{1}$ is calculated event by event by  the following equation:
\vspace{0.1cm}
\begin{equation}
\begin{aligned}
& \Psi_{1}^{\prime} = \Psi_{1} + \Delta \Psi_{1} \\
& \Delta \Psi_{1} = \sum_{i=1}^{20} \frac{2}{i}\left[-\left\langle\sin \left(i \Psi_{1}\right)\right\rangle \cos \left(i \Psi_{1}\right)+\left\langle\cos \left(i \Psi_{1}\right)\right\rangle \sin \left(i \Psi_{1}\right)\right]
\end{aligned}
  \label{Eq:Shift_psi}
\end{equation}

where the brackets refer to an average over events which are in the same centrality bins. 
$\Psi_{1}$ is the position weight corrected event plane azimuth and $\Psi_{1}^{\prime}$ is the event plane angel with shift calibration. 
After the shift calibration, a flat event plane distribution is achieved as shown by blue line in Fig.~\ref{Fig:EP_dis}, 
and the resolution between the left side and right side is consistent as shown in Fig.~\ref{Fig:whole_res}(c).

\begin{figure}[htb]
  \centering
  \includegraphics[width=3.6in,keepaspectratio]{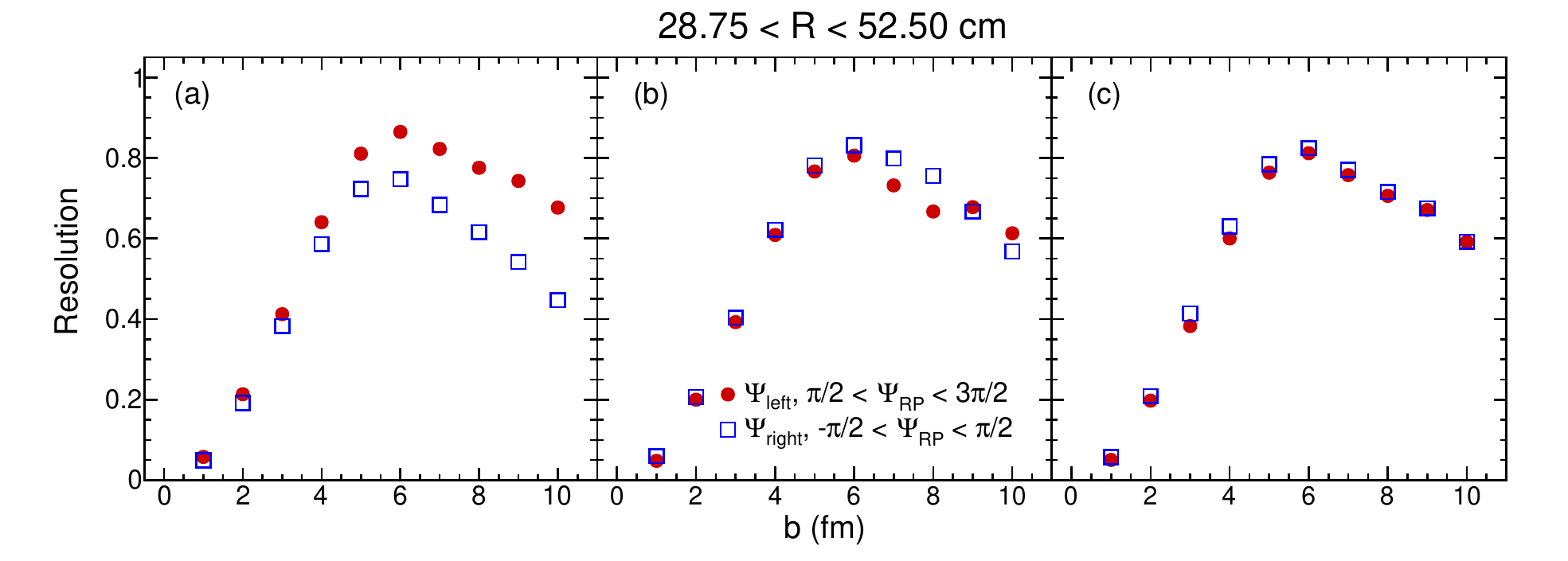}
  \caption{(a) The resolution of $1^{\rm st}$ order event plane from ZDC sub-ring radius $52.5 < R < 76.25$ cm without position weight.
  (b) The resolution of $1^{\rm st}$ order event plane from ZDC sub-ring radius $52.5 < R < 76.25$ cm with position weight.
  (c) The resolution of $1^{\rm st}$ order event plane from ZDC sub-ring radius $52.5 < R < 76.25$ cm with position weight and shift correction.}     
  \label{Fig:Sub_res}
\end{figure}

In experiment, event plane calculated from different rapidity windows helps us to understand the systematic uncertainties of flow measurements.
Correspondingly, the event plane from ZDC sub-rings which correspond to different rapidity windows is studied. 
Figure ~\ref{Fig:Sub_res} shows the $1^{\rm st}$ order event plane resolution from ZDC sub-ring 
radius $52.5 < R < 76.25$ cm without position weight~\ref{Fig:Sub_res}(a), with position weight~\ref{Fig:Sub_res}(b) and with position weight and shift correction~\ref{Fig:Sub_res}(c). 
These results indicate position weight and shift method also work well for event plane calculated by ZDC sub-ring.

After eliminating the resolution difference due to the asymmetric acceptance via position weight and shift method,
the $1^{\rm st}$ order event plane resolution from ZDC is calculated by using the two sub-event plane method~\cite{Voloshin:2008dg}. 
The full event is divided randomly into two independent sub-events with equal tracks, and the event plane resolution estimated by correlating two sub-events
as defined by Equ.~\ref{Eq:Sub_res}:
\begin{equation}
\begin{aligned}
R_{1, \mathrm{sub} }&= \sqrt{\left\langle\cos \left(\Psi_1^A-\Psi_1^B\right)\right\rangle}\\
&= \sqrt{\pi} / 2 \chi e^{(-\chi^2 / 2)} \left(I_{0} \left(\chi^2 / 2\right)+I_{1}\left(\chi^2 / 2\right)\right)
\end{aligned}
  \label{Eq:Sub_res}
\end{equation}
where A and B denote the two sub-events. 
Since the $\chi$ is proportional to square root of multiplicity and full event with twice particles as sub-events, 
the full event plane resolution is obtained by
\begin{equation}
R_{\text {full }}=R \left(\sqrt{2} \chi_{\text {sub }}\right)
  \label{Eq:Full_res}
\end{equation}

\begin{figure}[htb]
    \centering
    \includegraphics[width=2.5in,keepaspectratio]{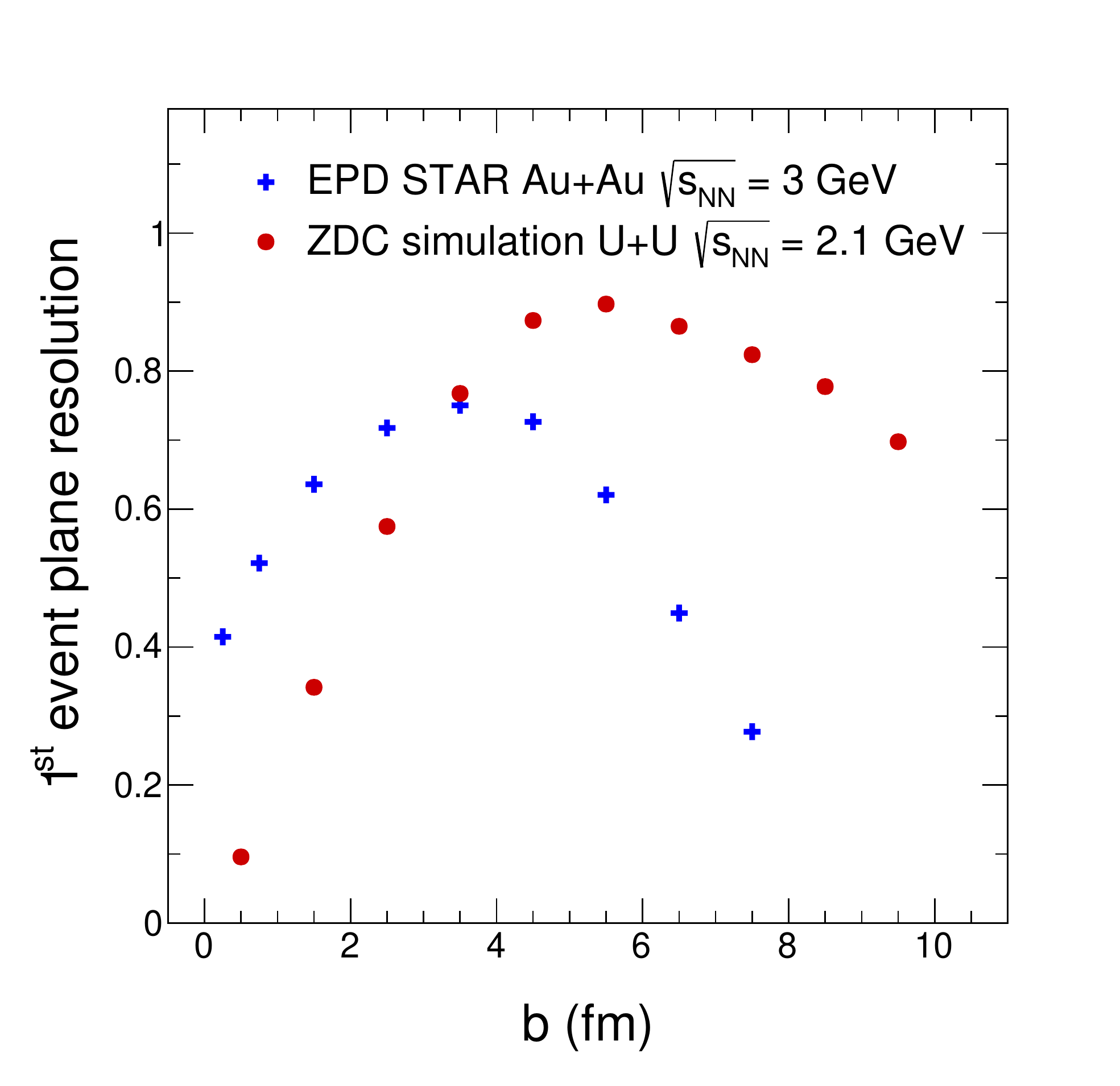}
    \caption{The resolution of $1^{\rm st}$ order event plane as a function of
    impact parameter from ZDC whole ring, comparing with $1^{\rm st}$ order event plane resolution from STAR EPD in Au+Au collisions at {\rootsNN} = 3.0 GeV.}
    \label{Fig:Final_res}
\end{figure}

The resolution of $1^{\rm st}$ order event plane as a function of impact parameter from ZDC whole ring, 
comparing with $1^{\rm st}$ order event plane resolution from STAR Event Plane Detector in Au+Au collisions at {\rootsNN} = 3.0 GeV~\cite{STAR:2021yiu} is shown in Fig.~\ref{Fig:Final_res}.
The event plane resolution from CEE-ZDC reaches $\sim 90\%$ in middle central collisions ($4 < b < 7$ fm). 
The resolution of ZDC is better in the region of $b < 4$ fm, but worse for $b > 4$ fm, which is probably due to the different sizes of gold and uranium nuclei, experimental acceptance and detector performance.

\begin{figure}[htb]
  \centering
  \includegraphics[width=3.65in,keepaspectratio]{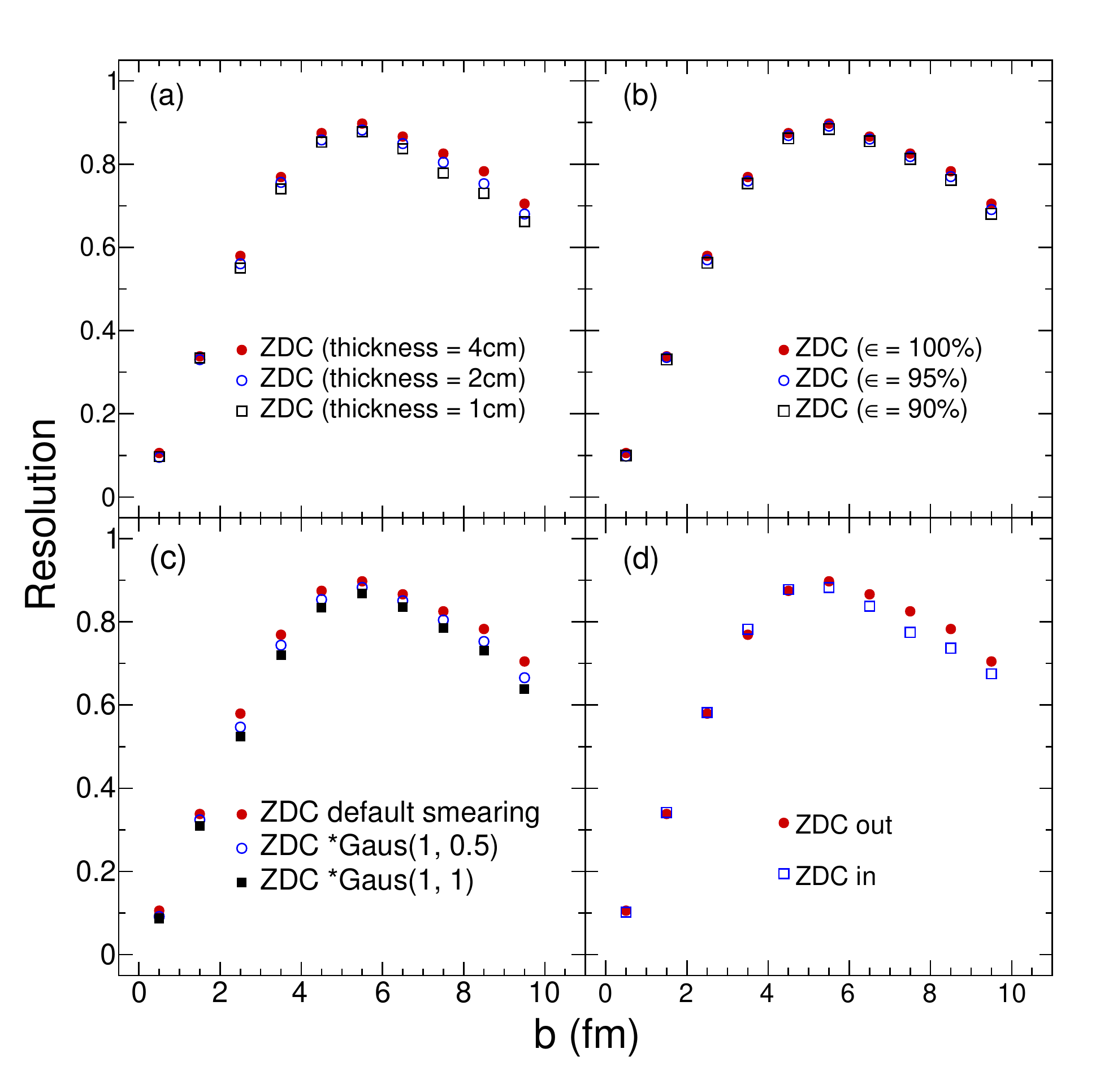}
  \caption{(a) The effect of ZDC thickness on the $1^{\rm st}$ order event plane resolution.
  (b)The effect of ZDC hit efficiency on the $1^{\rm st}$ order event plane resolution. 
  (c)The effect of ZDC energy resolution on the $1^{\rm st}$ order event plane resolution.
  (d)The effect of heavy nuclei de-excitation on the $1^{\rm st}$ order event plane resolution.}  
  \label{Fig:Sys_res}
\end{figure}

We also systematically investigate the effects of ZDC detector thickness, hit efficiency, energy resolution, and model dependence on the first-order event plane resolution.
 As shown in Fig.~\ref{Fig:Sys_res}, where the solid red dots represent the default version: ZDC thickness of 4 cm, hit efficiency of 100\%, default energy smearing as Equ.~\ref{Eq:Energy_smear} , and heavy nuclei from IQMD generator de-excitation. The effect of different variables is investigated one by one.
The resolution of $1^{\rm st}$ order event plane slightly decreases as the ZDC detector thickness decreases, as shown in Fig.~\ref{Fig:Sys_res}(a). It is because that more accurate measurement of deposited energy is archived by a thicker ZDC.  
Fig.~\ref{Fig:Sys_res}(b) shows the hit efficiency dependence of $1^{\rm st}$ order event plane resolution.
The ZDC hit efficiency is reduced to 90\%, and the event plane resolution is almost unchanged.
The effect of ZDC energy resolution is investigated by applying an additional Gaussian smearing to the deposited energy, where Gaussian(1, 0.5) is with center value 1 and width 0.5, and Gaussian(1, 1) is with center value 1 and width 1, respectively. The smaller Gaussian width represents better energy resolution. As the energy resolution decreases, the first-order event plane resolution decreases by about 5-10\% as shown in Fig.~\ref{Fig:Sys_res}(c).
Figure~\ref{Fig:Sys_res}(d) shows the relationship between the ZDC event plane resolution and the IQMD heavy nuclei de-excitation, where $"$out$"$/$"$in$"$ means the heavy nuclei are de-excitation or not. 
The resolution estimated with the IQMD model with heavy nuclei de-excitation is slight higher than IQMD without heavy nuclei de-excitation, as the multiplicity is higher in the former case.

\begin{figure*}[!htbp]
    \centering
    \includegraphics[width=0.9\textwidth]{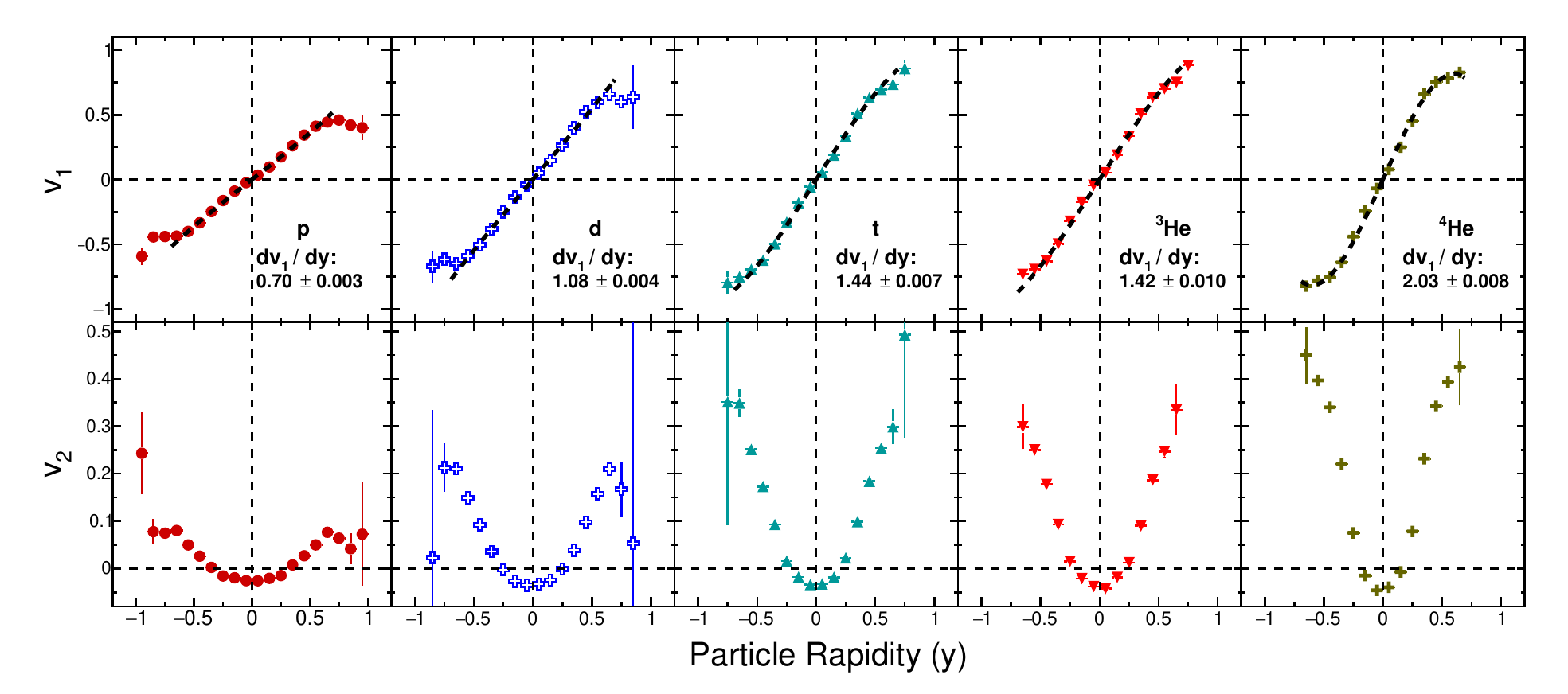}
    \caption{$v_{1}$ and $v_{2}$ as a function of rapidity for protons, deuteronss, tritons, $^3$He, $^4$He with collision impact parameter $1< b < 4$ fm from IQMD $^{238}$U + $^{238}$U 500 MeV/u ({\rootsNN} = 2.1 GeV). The $v_1$ slopes are extracted by: $y=ax+bx^{3}$.}
    \label{Fig:v1v2_results}
\end{figure*}

\section{Collectivity flow predictions from IQMD model}\label{sec.IV}

Collective flow is sensitive to the details of the expansion of the produced medium during the early collision stage.
Flow measurements at CEE would offer information of the QCD phase structure at high baryon density region.
Collectivity flow predictions are presented at a typical CEE collision energy based on IQMD model. 
Figure~\ref{Fig:v1v2_results} shows the $v_{1}$ and $v_{2}$ as a function of rapidity for 
protons, deutons, tritons, $^3$He, $^4$He with impact parameter $1< b < 4$ fm 
from IQMD $^{238}$U + $^{238}$U collisions at 500 MeV/u ({\rootsNN} = 2.1 GeV).
The $v_1$ slope values extracted by: $y=ax+bx^{3}$ strongly depends on the nuclei number. 
The $v_2$ values are negative in the middle rapidity due to the squeeze-out effect--the medium expansion is shadowed by spectator nucleons, particles are preferred to emit in the direction perpendicular to reaction plane~\cite{Voloshin:2008dg}, while $v_2$ becomes positive in the forward rapidity as the squeeze-out effect becomes weak. Similar as $v_1$ slope, $v_2$ values also show a strong dependence on the nuclei number.

\begin{figure}[htbp]
    \centering
    \includegraphics[width=2.5in,keepaspectratio]{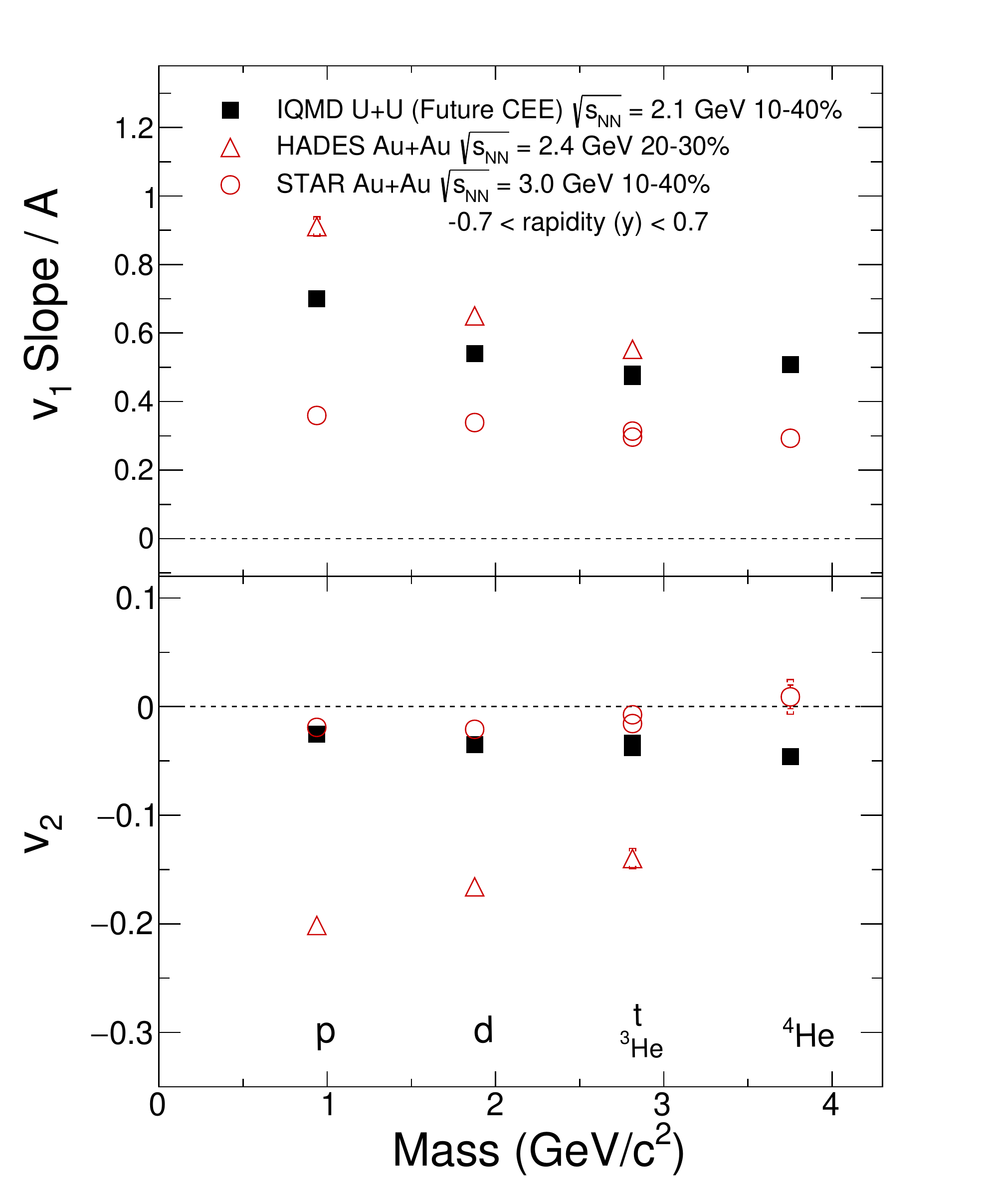}
    \caption{Atomic number $A$ scaled $v_{1}$ slope (upper panel) and $v_2$ (lower panel) at middle rapidity for protons, deuterons, tritons, $^3$He, $^4$He from HADES, STAR experiments and IQMD model calculations for CEE. The brackets represent the systematic uncertainties on the experimental data.}
    \label{Fig:vn_mass}
\end{figure}

Figure~\ref{Fig:vn_mass} presents 
the $d(v_{1}/A)/dy$ and $v_{2}$ for protons, deuterons, tritons, $^3$He, $^4$He from HADES~\cite{HADES:2020lob} and STAR~\cite{STAR:3GeV_light} experiments together with IQMD model calculations $\footnote{Unlike in the experiment, the centrality here is determined from the impact parameter in the model calculations.}$,
where $A$ is the atomic number. $v_{1}/A$ represents the directed flow carried by each nucleon in light nuclei, and the scaling behavior suggests the coalescence production mechanism of light nuclei in the heavy-ion collisions.
$v_{2}$ is calculated in the rapidity range of $-0.1 < y < 0$ for STAR experiment and IQMD model calculations, and $-0.05 < y < 0.05$ for HADES experiment respectively.
The atomic number scaled $v_1$ slope from HADES and IQMD shows a decrease trend with increase of 
atomic number, while STAR data weakly depends on the atomic number in collisions at {\rootsNN} = 3 GeV.
The absolute value of $v_{2}$ from HADES decreases with increase of atomic number, while the results for STAR and IQMD are almost unchanged with atomic number.
It may indicate the light nuclei is not purely formed by coalescence mechanism in Au+Au collisions at 
{\rootsNN} = 2.4 GeV, 
while coalescence is the dominant production mechanism in Au+Au {\rootsNN} = 3.0 GeV.
The production of light nuclei in the IQMD model is a mixture of light nuclei fragments and coalescence of nucleons and light nuclei.
The dominance of production mechanism in the IQMD model depends on the collision energy and parameter settings.
The predictions given by IQMD model in U+U collisions at {\rootsNN} = 2.1 GeV will be validated in future CEE experiments.

Future measurement of $v_1$ and $v_2$ will help us to study the Equation of State of the produced nuclear matter at CEE energies~\cite{Russotto:2013fza, Wang:2022det}, as well as understand the production mechanism of light nuclei in the high baryon density region~\cite{Zhang:2009ba, Steinheimer:2012tb, Wang:2019eec, Yan:2006bx, Fang:2023sna}.

\section{Summary} \label{sec.V}

In this paper, we illustrate the procedures of event plane determination from ZDC at CEE. 
The calculations from IQMD Monte-Carlo event generator (500 MeV/u $^{238}$U + $^{238}$U) 
are used as inputs and the detector environment is simulated by GEANT4. 
In order to correct the bias caused by dipole magnet, a position dependent weight is introduced to calibrate the asymmetric acceptance. 
After an additional shift correction, the resulting first order event plane resolution reaches as high as $\sim 90\%$ in middle central collisions ($4 < b < 7$ fm).
Collective flow $v_{1}$ and $v_{2}$, as a function of rapidity, for $p$, $d$, $t$, $^3$He and $^4$He in middle central collisions are presented based on the IQMD model. These results are compared with experimental data from 2.4 GeV and 3 GeV Au+Au collisions at HADES and STAR experiment, respectively. 
The measurements from HADES and STAR experiments suggest the coalescence is the dominant production mechanism of light nuclei at 3 GeV, while light nuclei fragments and coalescence are both important at 2.4 GeV. The predictions from IQMD at 2.1 GeV will be validated in future CEE experiments.

\section*{Acknowledgments}
We thank Prof. Li Ou and Zhigang Xiao for generating IQMD data and fruitful discussions.


\end{document}